\def   \ni {\noindent}

\def   \ssk {\vskip  5truept}

\def   \bsk {\vskip 15truept}

\def   \newline {\hfil\break}

\documentstyle[epsfig]{article}
\begin{document}

\hsize 5truein
\vsize 8truein
\font\abstract=cmr8
\font\keywords=cmr8
\font\caption=cmr8
\font\references=cmr8
\font\text=cmr10
\font\affiliation=cmssi10
\font\author=cmss10
\font\mc=cmss8
\font\title=cmssbx10 scaled\magstep2
\font\alcit=cmti7 scaled\magstephalf
\font\alcin=cmr6 
\font\ita=cmti8
\font\mma=cmr8
\def\ref{\par\noindent\hangindent 15pt}
\null
%\vskip 3.0truecm
%\baselineskip = 12pt

% beginning of font "title"
\title{\ni SAX J1810.8-2609: an X-ray bursting transient}                                               

% beginning of font "author and affiliation"
\bsk \bsk
\author{\ni M.~Cocchi $^{1}$, A.~Bazzano $^{1}$, L.~Natalucci $^{1}$, 
 P.~Ubertini $^{1}$, J.~Heise $^{2}$, J.M.~Muller $^{2,3}$, M.J.~Smith $^{2,3}$, 
 and J.J.M.~in~'t~Zand $^{2}$}                                                       
\bsk
\affiliation{
 1) Istituto di Astrofisica Spaziale (IAS/CNR ), via Fosso del Cavaliere, 00133 Roma, Italy.}
\affiliation{
 2) Space Research Organisation Netherlands (SRON), Sorbonnelaan 2, 3584 CA, Utrecht, the Netherlands.}
\affiliation{
 3) {\em BeppoSAX} Science Data Centre, Nuova Telespazio, via Corcolle 19, 00131 Roma, Italy.}
\bsk
\baselineskip = 12pt

% beginning of font "abstract and keywords"
\abstract{ABSTRACT\\
\ni
We report on the {\em BeppoSAX}-WFC observation of a strong type-I X-ray burst from the recently discovered transient source 
{\em SAX J1810.8-2609} (Ubertini et al., 1998a).  The observed event showed evidence of photospheric radius 
expansion due to super-Eddington burst luminosity.  Burst detection identifies the source as a probable 
low-mass X-ray binary harbouring a neutron star.
}                                                    
\bsk
\baselineskip = 12pt

\bsk
\baselineskip = 12pt

% beginning of font "text"

\text{\ni 1. INTRODUCTION \\
{\em SAX~J1810.8-2609} was observed for the first time in the 2-28 keV band by the Wide Field Cameras (WFC) on board the 
{\em BeppoSAX} satellite (Ubertini et al., 1998a) during a long term monitoring campaign of the Galactic Bulge region.  This 
weak transient was detected on 1998 March 10 at an intensity of 15 mCrab in the 2-9 keV band at the 
position $\alpha=18^{\rm h}10^{\rm m}46^{\rm s}$, $\delta=-26^{\circ}09^{\prime}.1$ (J2000), error radius $3^{\prime}$.
During the observation a strong burst (peak intensity of $\sim1.6$ Crab) lasting $\sim45$~s was detected from a position consistent 
with that of the persistent emission. \\
Not surprisingly, {\em SAX~J1810.8-2609} was not observed by {\em RXTE}-ASM, being the source flux only $\sim 1$ ASM count/s at its 
maximum intensity. \\
The source was monitored on March 12 during a follow-up observation with the {\em BeppoSAX} Narrow-Field Instruments 
(Ubertini et al., 1998b) .   At that time the 2-10 keV source intensity had declined to $\sim 7.4$ mCrab, and the spectrum was 
consistent with a power-law model with photon index $\Gamma \simeq -2.2$. \\
Another follow-up observation was performed on March 24 with the HRI on board {\em ROSAT} (Greiner et al., 1998).  A low 
energy source was found at a position consistent with the WFC one (coordinates: $\alpha=18^{\rm h}10^{\rm m}44^{\rm s}.5$, 
$\delta=-26^{\circ} 09^{\prime}01^{\prime\prime}$, error radius $10^{\prime\prime}$).  
The 0.1-2.4 keV flux was $\sim 1.5$ mCrab. Several optical objects are within the HRI error box of the source. \\
{\em SAX J1810.8-2609} was not detected in previous {\em ROSAT} observations of the same region on 1993 September 10 (0.1-2.4 
keV $3\sigma$ upper limit of $\sim 0.08$ mCrab) and in 1990 during the All-Sky Survey, thus indicating the transient nature of the 
source.
\ssk
\ni

%--------------------------  figure 1
%this section shows how to insert a figure in the text
\begin{figure} [h]
\centerline{\psfig{file=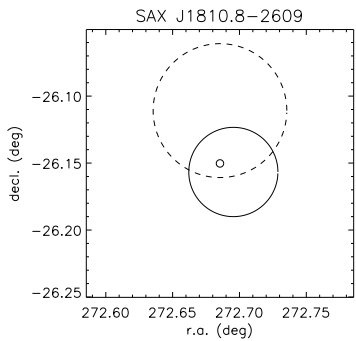, width=5cm}}
\caption{FIGURE 1. A sky map showing the {\em BeppoSAX} and {\em ROSAT} error regions.\\
  small solid circle: {\em ROSAT} error circle (Greiner et al. 1998);~~large solid circle: {\em BeppoSAX}-WFC error 
  circle;~~large dashed circle: revised ($3^{\prime}$ radius) {\em BeppoSAX}-NFI error circle.
}
\end{figure}
%---------------------------------
%--------------------------  figure 2
%this section shows how to insert a figure in the text
\begin{figure} [h]
\centerline{\psfig{file=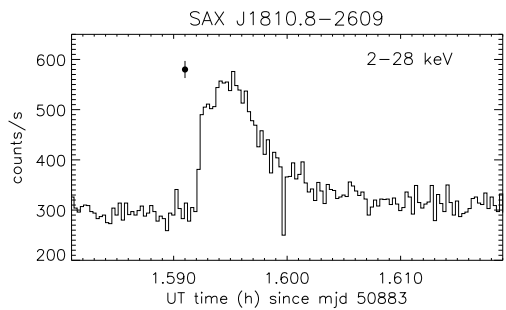, width=7.4cm}}
\caption{FIGURE 2. The 2-28 keV time history of the burst (1~s time bin).
}
\end{figure}
%---------------------------------

\bsk
\ni 2. OBSERVATION AND DATA ANALYSIS \\
The Wide Field Cameras (WFC) on board {\em BeppoSAX} satellite consist of two identical coded mask telescopes (Jager et al., 
1997).  The two cameras point at opposite directions each covering a $40^{\circ}\times 40^{\circ}$ field of view. 
With their source location 
accuracy of $0^{\prime}.6$ (68\% confidence level), a time resolution of 0.488 ms, and an energy resolution of 18\% at 6 keV, 
the WFCs are very effective in studying hard X-ray (2-28 keV) transient phenomena.  The imaging capability, combined 
with the good instrument sensitivity (a few mCrab in $10^{4}{\rm s}$), allows accurate monitoring of complex sky regions like the 
Galactic Bulge. \\
The data of the two cameras are systematically searched for bursts or flares by analysing the time profiles of the detector in 
the 2-11 keV energy range with a 1~s time resolution.  Reconstructed sky images are generated in coincidence with any 
statistically meaningful event, to identify possible bursters.  The accuracy of the reconstructed position, which of 
course depends on the intensity of the burst, is typically better than $5^{\prime}$.  
This analysis procedure allowed the identification of $\sim 630$ X-ray bursts in a total of about 1.43 Ms WFC effective  
observing time (see e.g. Cocchi et al., 1998).

%--------------------------  figure 3
%this section shows how to insert a figure in the text
\begin{figure} [h]
\centerline{\psfig{file=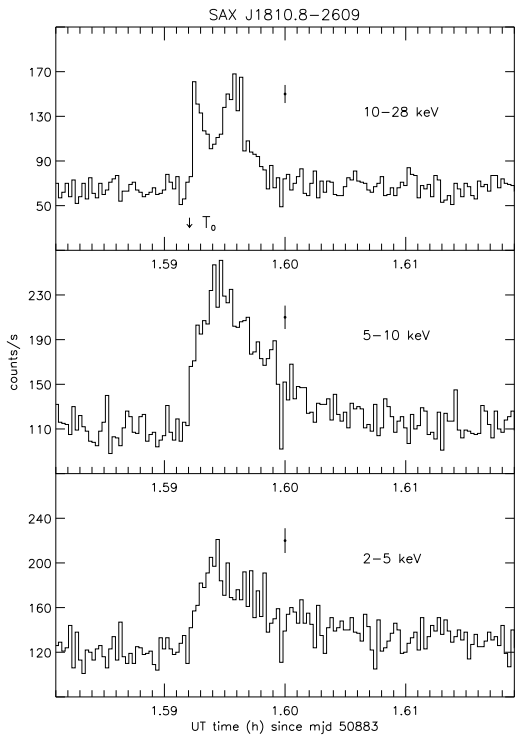, width=7.4cm}}
\caption{FIGURE 3. The burst profile in 3 different energy bands (1~s time bin). 
 The arrow in the upper panel indicates $T_{0}$ (see Table 1).
}
\end{figure}
%---------------------------------

The March 10.06633 UT burst of {\em SAX J1810.8-2609} is up to now the only X-ray burst observed from this transient source.   
The WFC determination of the source position was improved with respect to the preliminary
one reported by Ubertini et al., 1998a. The coordinates are $\alpha=18^{\rm h}10^{\rm m}46.9^{\rm s}$, 
$\delta=-26^{\circ}09^{\prime}24^{\prime\prime}$ ($2^{\prime}$ error radius), confirming the association with the {\em ROSAT} 
source (see Figure 1).\\
The burst had a peak intensity of $1.89~\pm~0.18$ Crab in the 2-28 keV band and its 
time history is shown in Figure 2.  The profile looks rather typical, being FRED-shaped with an  
$e$-folding time of $12.5\pm0.7$~s.   But energy resolved time profiles (Figure 3) show a double-peaked structure at 
high energy (10-28 keV) suggesting photospheric radius expansion during a very bright (super-Eddington) type-I burst 
(Lewin et al., 1995, and references therein). \\
The overall spectrum of the burst as obtained by accumulating 48~s of data with the best S/N ratio can be modelled by a thermal 
blackbody emission with colour temperature ${\em k}T \simeq 2$ keV (Table 1).  The $N_{\rm H}$ value cannot be satisfactorily 
constrained, since we obtain $(1.6 \pm 0.7)\times 10^{22}~{\rm H~cm}^{-2}$.\\
Time resolved spectra of the burst were calculated in coincidence with the peak structures observed in the high energy 
profile.  The fit parameters and the $R_{\rm km}/d_{\rm 10 kpc}$ values are consistent with a radius expansion of a factor of $\sim 2$
during the first $\sim10$~s of the event (see Table 1).
After the subsequent contraction of the emitting region, the typical spectral softening due to the cooling of the photosphere is 
observed. \\

\begin{table*} [h] %hbt
% space before first and after last column: 1.5pc
% space between columns: 3.0pc (twice the above)
\setlength{\tabcolsep}{0.3pc}
% -----------------------------------------------------
% adapted from TeX book, p. 241
\newlength{\digitwidth} \settowidth{\digitwidth}{\rm 0}
\catcode`?=\active \def?{\kern\digitwidth}
% -----------------------------------------------------
\caption{Table~1: Summary~of~the~results~of~the~time~resolved~spectral~analysis~of~the~burst}
\label{tab:SAX 1810 bursts}
\begin{tabular*}{\textwidth}{@{}l@{\extracolsep{\fill}}rrrr}
\hline
		   \multicolumn{1}{r}{Data set}
                 & \multicolumn{1}{r}{Time range(*)} 
                 & \multicolumn{1}{r}{Blackbody {\em k}T (keV)}
                 & \multicolumn{1}{r}{$    \chi^{2}_{\rm r}$ (26 dof)}
                 & \multicolumn{1}{r}{$R_{\rm km}/d_{\rm 10 kpc}$}
                          \\
\hline
Burst average    & $T_{0} \div T_{0}+48s$        &  $1.98 \pm 0.04$  &  $1.89$  &  $12.3 \pm 0.6$ \\
Peak~1         & $T_{0}+1s \div T_{0}+5s$   &  $2.55 \pm 0.11$  &  $1.46$  &  $ 9.5 \pm 1.1$ \\
Tail~1         & $T_{0}+5s \div T_{0}+10s$  &  $1.83 \pm 0.06$  &  $1.29$  &  $20.1 \pm 1.7$ \\
Peak~2         & $T_{0}+10s \div T_{0}+16s$ &  $2.71 \pm 0.09$  &  $0.98$  &  $ 9.72 \pm 0.83$ \\
Tail~2         & $T_{0}+16s \div T_{0}+23s$ &  $1.94 \pm 0.08$  &  $1.09$  &  $14.6 \pm 1.6$ \\
Tail~3         & $T_{0}+23s \div T_{0}+48s$ &  $1.53 \pm 0.07$  &  $1.57$  &  $13.2 \pm 1.7$ \\
\hline
\multicolumn{5}{@{}p{120mm}}{(*) $T_{0} = 1.59206$ UT}
\end{tabular*}
\end{table*}

The data analysis results indicate the transient source {\em SAX J1810.8-2609} as a type-I X-ray burster and, most likely, as a low-mass 
X-ray binary containing a weakly magnetised neutron star.

Eddington-luminosity X-ray bursts can lead to an estimate of the source distance.  Taking into account the observed 2-28 
keV peak flux of 1.89 Crab and assuming a $2\times 10^{38} {\rm erg/s}$ Eddington bolometric luminosity for a $1.4 {\rm M}_{\odot}$ 
neutron star, we obtain d = $4.9~\pm~0.3$ kpc.  
%If we assume the average luminosity of super-Eddington bursts proposed by
%Lewin et al., 1995 ($3.0 \pm 0.6 \times 10^{38} {\rm erg/s}$) we obtain d = $6.0~\pm~0.9$ kpc.
We derive a steady source bolometric luminosity of $\sim 3\times 10^{36}{\rm erg/s}$ at the 
epoch of its discovery (1998 March 10).  We also obtain an average radius of $\sim 6$ km for the blackbody emitting region 
during the burst, a value supporting the neutron-star nature of the collapsed object.
\ssk
\ni

}

\bsk
\baselineskip = 12pt
{\abstract \ni ACKNOWLEDGMENTS \\
We thank the staff of the BeppoSAX Science Operation Centre and Science
Data Centre for their help in carrying out and processing the WFC Galactic Centre
observations. The BeppoSAX satellite is a joint Italian and Dutch program.
M.C., A.B., L.N. and P.U. thank Agenzia Spaziale Nazionale (ASI) for grant support.
}

\bsk
\baselineskip = 12pt

% beginning of font "references"

{\references \ni REFERENCES
\ssk
\ref Cocchi, M., et al. 1998, Nucl.Phys. B (Proc. Suppl.) 69/1-3, 232
\ref Greiner, J., Castro-Tirado, A.J., and Boller, T. 1998, IAUC 6985
\ref Jager R., et al. 1997, A\&A, 125, 557
\ref Lewin, W.H.G., van Paradijs, J., and Taam, R.E. 1995, in "$"$X-ray Binaries$"$", ed. W. Lewin, J. van Paradijs 
		\& E. van den Heuvel, Cambridge University Press, Cambridge, p. 175
\ref Ubertini, P., et al. 1998a, IAUC 6838
\ref Ubertini, P., et al. 1998b, IAUC 6843
}

%--------------------------  figure 3
%this section shows how to insert a figure in the text
%%\begin{figure} [h]
%%\centerline{\psfig{file=s18h2plt.ps, width=12cm}}
%%\caption{FIGURE 3. The spectrum in the P-2 time range.
%%}
%\end{figure}
%---------------------------------
%--------------------------  figure 4
%this section shows how to insert a figure in the text
%%\begin{figure} [h]
%%\centerline{\psfig{file=h_ratio.ps, width=11cm}}
%%\caption{FIGURE 4. Time history of the 10-28 keV / 2-10 keV hardness ratio during the burst.
%%}
%%\end{figure}
%---------------------------------

\end{document}